\documentclass[twocolumn,showpacs,amsmath,amssymb,prb,superscriptaddress]{revtex4-1}
\usepackage{graphicx}
\usepackage{color}

\begin{document}

\title{Universal Frequency Dependence of the Hopping AC Conductance
in p-Ge/GeSi Structures in the Integer Quantum Hall Effect Regime}

\author{I.~L.~Drichko}
\affiliation{Ioffe Institute, 26 Politekhnicheskaya, 194021 St.~Petersburg, Russia}

\author{A.~A.~Dmitriev}
\affiliation{Ioffe Institute, 26 Politekhnicheskaya, 194021 St.~Petersburg, Russia}
\affiliation{Department of~Nanophotonics and Metamaterials, ITMO University, 49 Kronverksky Pr., 197101 St.~Petersburg, Russia}

\author{V.~A.~Malysh}
\affiliation{Ioffe Institute, 26 Politekhnicheskaya, 194021 St.~Petersburg, Russia}

\author{I.~Yu.~Smirnov}
\affiliation{Ioffe Institute, 26 Politekhnicheskaya, 194021 St.~Petersburg, Russia}

\author{Y.~M.~Galperin}
\affiliation{Department of Physics, University of Oslo, 0316 Oslo, Norway}
\affiliation{Ioffe Institute, 26 Politekhnicheskaya, 194021 St.~Petersburg, Russia}

\author{H.~von~K{\"a}nel}
\affiliation{Laboratory for Solid State Physics, ETH Zurich, Otto-Stern-Weg 1, CH-8093 Zurich, Switzerland}

\author{M.~Kummer}
\affiliation{Laboratory for Solid State Physics, ETH Zurich, Otto-Stern-Weg 1, CH-8093 Zurich, Switzerland}

\author{D.~Chrastina}
\affiliation{L-NESS, Dipartimento di~Fisica, Politecnico di~Milano, Polo Regionale di~Como, Via Anzani 52, I-22100 Como, Italy}

\author{G.~Isella}
\affiliation{L-NESS, Dipartimento di~Fisica, Politecnico di~Milano, Polo Regionale di~Como, Via Anzani 52, I-22100 Como, Italy}

\date{\today}

\begin{abstract}
The hopping ac conductance, which is realized at the transverse conductance minima in the regime of the integer Hall effect, has been measured using a combination of acoustic and microwave methods. Measurements have been made in the p-GeSi/Ge/GeSi structures with quantum wells in a wide frequency range (30$\div$1200~MHz). The experimental frequency dependences of the real part of ac conductance $\sigma_1$ have been interpreted on the basis of the model presuming hops between localized electronic states belonging to isolated clusters. At high frequencies, dominating clusters are pairs of close states; upon a decrease in frequency, large clusters that merge into an infinite percolation cluster as the frequency tends to zero become important. In this case, the frequency dependences of the ac conductance can be represented by a universal curve. The scaling parameters and their magnetic-field dependence have been determined.
\end{abstract}


\maketitle

\section{Introduction}

Low-dimensional electronic systems have been
objects of intense investigations for many years. The
2D electron gas in a quantizing transverse magnetic
field is an informative object that exhibits different
electric conduction mechanisms depending on the
Landau level filling factors. The most known
regimes include the integer and fractional quantum
Hall effects~\cite{QHE} and the Wigner crystal. In the whole, a
rich physical pattern is formed with details depending
on the competition between the effects of structural
disorder and the electron–electron interaction.

In this work, we are studying the hopping ac conductance
of a 2D gas of holes, which is realized in
$\text{p-Ge/GeSi}$ structures at the minima of the integer
quantum Hall effect (IQHE). In these regions, the
electronic states are localized by disorder, and conduction
occurs as hops between localized states. It is
interesting that the physical patterns of the dc and ac
conduction are substantially different in such a
regime. In the former case, the hops occur over the
states of the so-called percolation cluster that pierces
the entire sample and terminates at the electrodes~\cite{Shklovskii2}.
In the latter case, hopping occurs over the states of
finite clusters isolated from one another (see, for
example, review~\cite{Parshin}). Therefore, comparative analysis of dc and ac conductions provides additional information.

\begin{figure}[t]
\centerline{
\includegraphics[width=7.9cm,clip=]{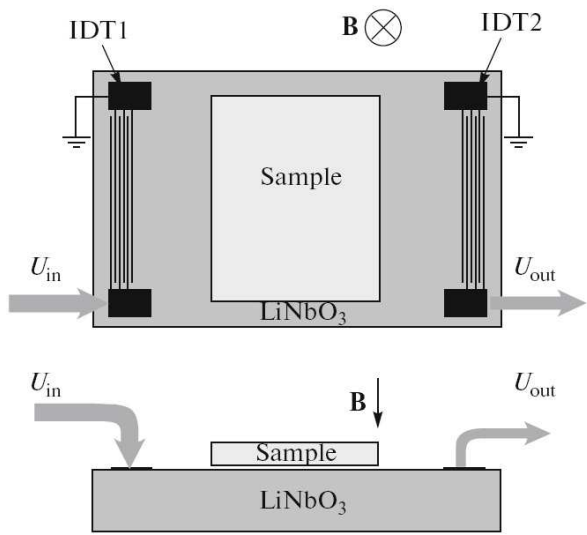}
} \caption{Illustration the acoustic method. A SAW is
generated by interdigital transducer 1 (IDT1) and is
detected by IDT2 after the passage over the surface of the
LiNbO3 piezoelectric substrate.
\label{Fig1_X2017scaling}}
\end{figure}

The ac transport is studied using various experimental
techniques among which contact-free methods
occupy a special place. One of them is the acoustic
method~\cite{Wixforth1989} based on the Rayleigh-type surface
acoustic wave (SAW) propagating over the surface of
a LiNbO3 piezoelectric substrate, to which the experimental
sample is pressed (Fig.~\ref{Fig1_X2017scaling}). In this case, a
deformation wave in the substrate is accompanied
with an electric field varying with the SAW frequency,
which penetrates into the sample and interacts with
charge carriers. As a result of such an interaction, the
SAW intensity decreases (absorption) and the SAW
velocity changes, which is registered by the detecting
system. Then the complex ac conductance of the
investigated sample is calculated from the experimentally
measured SAW absorption coefficient and the
change in the SAW velocity. It should be noted that in
this method, the electric field polarization vector in
the sample is parallel to the SAW wavevector, and the
SAW itself is used only for introducing an electric field
into the experimental sample by the contactless technique
without deforming it.

A certain disadvantage of the acoustic method is
the limited range of working frequencies, which is mainly determined by the configuration of interdigital
transducers used for generating and detecting SAWs.
Since analysis of the frequency dependence of the
conductance makes it possible to study the mechanisms
of localization of charge carriers in quantum size
systems,~\cite{Drichko2000,Drichko2005,DrichkoPhysicaE} it is important to perform measurements
in a wide frequency range.

\begin{figure}[t]
\centerline{
\includegraphics[width=7.9cm,clip=]{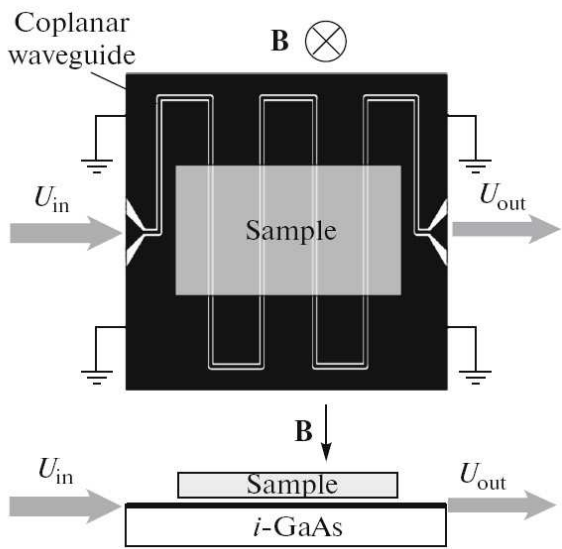}
} \caption{Sketch of the microwave technique.
\label{Fig2_X2017scaling}}
\end{figure}

Such a possibility is ensured by the microwave contactless
technique~\cite{Engel1993} based on the use of a quasitransverse
electromagnetic wave (quasi-TEM wave)
propagating in a coplanar waveguide formed on the
surface of insulating GaAs substrate (Fig.~\ref{Fig2_X2017scaling}). The
coplanar waveguide is prepared in the form of a meander
for increasing the length over which charge carriers
interact with the electric field of a quasi-TEM wave in
the sample under investigation.

In this case, the experimental sample is pressed to
the coplanar waveguide. Analogously to the electric
field of the SAW in the acoustic method, the electric
field of the quasi-TEM wave penetrates into the sample
and interacts with charge carriers; as a result, the
wave amplitude decreases and its phase changes.
These quantities are subsequently used for determining
the ac conductance of the sample under investigation.~\cite{Endo} In this case, the electric field polarization
vector is directed in the same way as in the acoustic
method, and the electric field of the quasi-TEM wave
is also introduced into the sample by the contactless
method. The range of the working frequencies in the
microwave method is wider and is in fact determined
by the potentialities of the measuring instruments used
in the experiment and not by the configuration of the
coplanar waveguide itself. At the same time, the disadvantage
of the microwave technique is that the absolute
values of the ac conductance cannot be determined
in view of specific features of this method.

This study aims at analysis of the low-temperature
mechanisms of ac conduction in the Ge- and Si-based
objects in a wide frequency range. For this purpose,
the above-mentioned acoustic and microwave techniques
were used. We employed the method of comparison
of the results obtained using these two techniques,
which was proposed in~[\onlinecite{Drichko2014}]; in this method,
microwave measurements can be calibrated in the
absolute values of conductance by comparing with the
results of acoustic measurements in the accessible frequency
range. It will be shown below that such a calibration
must be performed in magnetic fields that correspond
to the frequency-independent conductance
maxima, which are determined by extended charge
carriers.

\section{Samples}
We have studied two
p-Ge/SiGe samples - 1
 with 13~nm quantum well with hole densities from $4 \times 10^{11}~\text{cm}^{-2}$, and 2 - with 20~nm quantum well with and $p=6 \times 10^{11}~\text{cm}^{-2}$, which were investigated
in detail in [\onlinecite{Drichko2014}]. These data will be used in the
discussion of experimental results.

\begin{figure}[t]
\centerline{
\includegraphics[width=7.9cm,clip=]{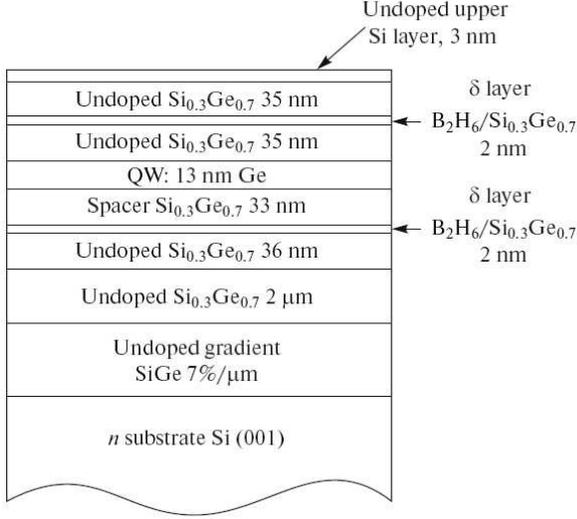}
} \caption{Structure of sample no. 1.
\label{Fig3_X2017scaling}}
\end{figure}

The structure of sample no. 1 is shown schematically
in Fig.~\ref{Fig3_X2017scaling}. It was obtained by chemical deposition
from the vapor phase with the help of a low-energy
plasma beam (LEPECVD).~\cite{Rosenblad} The active part of the
sample is a 2D channel in a stressed Ge layer (of thickness
13 nm), which is confined between SiGe layers
with a Ge concentration of about 70$\%$. Modulated
doping was performed by introducing diluted diborane
(B$_2$H$_6$) in the layers above and below the channel. The
concentration and mobility of holes in the 2D channel
at $T$=1.7~K were $p$=4$\times10^{11}$~cm$^{-2}$ and $\mu$=4.4$\times
10^4$~cm$^2$/(V$\cdot$s), respectively. The 2D Ge channel is stressed due to tensile strain under which the subband
of heavy holes turns out to correspond to a higher
energy. The energy gap between the subbands of light
and heavy holes is so wide that only heavy holes participate
in conduction in the entire accessible temperature
range.

\section{Experimental results}

The SAW absorption coefficient and the velocity
variations in sample no. 1 were measured in the temperature
interval (1.7–4.2) K in magnetic fields up to
8 T in the SAW frequency range from 30 to 200 MHz.

\begin{figure}[t]
\centerline{
\includegraphics[width=7.9cm,clip=]{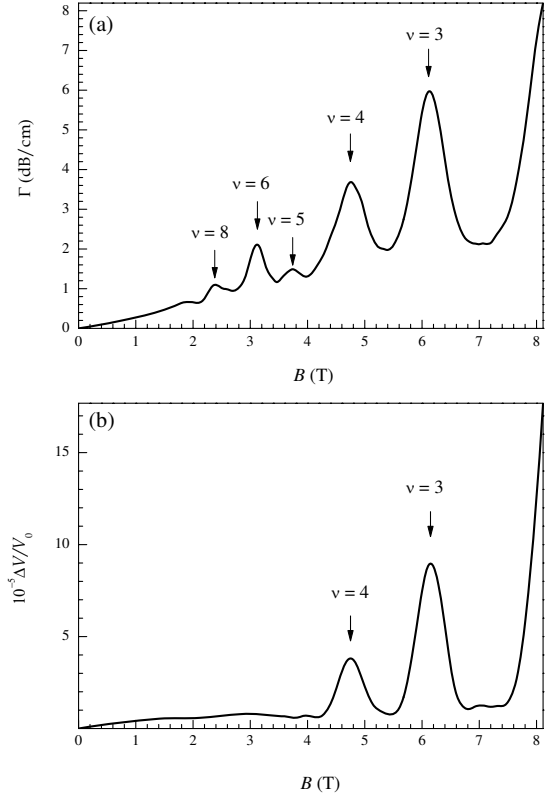}
} \caption{Dependences $\Gamma(B)$ (a), $\Delta V/V(B)$ (b),
$f$=142~MHz, $T$=4.2~K, $\nu$ is the filling factor
\label{Fig4_X2017scaling}}
\end{figure}

Figures \ref{Fig4_X2017scaling}a and \ref{Fig4_X2017scaling}b show the magnetic-field dependence
of the SAW absorption coefficient, $\Delta\Gamma(B)$=$\Gamma(B)$-$\Gamma(0)$$ \approx $$\Gamma(B)$ (since $\Gamma (0)$ $\ll$ $\Gamma (B)$)) and the magnetic-field dependence
of the SAW velocity variation, $\Delta
V(B)/V(0)$=[$V(B)$-$V(0)$]/$V(0)$ ($f$=142~MHz, $T$=4.2~K). It can be seen that oscillations
of the absorption coefficient and SAW velocity
appearing in the magnetic field correspond to Shubnikov–
de Haas oscillations below 3~T and to oscillations
in the regime of the integer quantum Hall effect
(IQHE) above 3~T. Analogous curves were obtained
for other frequencies and temperatures, as well as for
sample no. 2.

\begin{figure}[t]
\centerline{
\includegraphics[width=7.9cm,clip=]{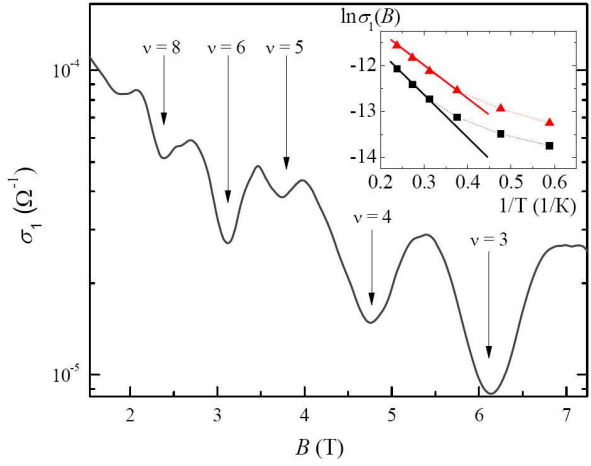}
} \caption{Dependence $\sigma_1 (B)$ for frequency $f$=142~MHz at
$T$=4.2~K; $\nu$ is the filling factor. The inset shows the
dependence of $\ln \sigma_1(B)$ on inverse temperature at oscillation
minima with filling factor $\nu$=3 (squares)
and 4 (triangles).
\label{Fig5_X2017scaling}}
\end{figure}

Dependences $\Gamma (B)$ and $\Delta V(B)/V(0)$ can be
expressed in terms of the real ($\sigma_1$) and imaginary ($\sigma_2$)
ac conductance components ($\sigma_{xx}^{AC}=\sigma_1-i\sigma_2$) using the
following expressions:~\cite{KaganSAW}

\begin{eqnarray}
  \label{eq:G}
&&\Gamma=8.68\frac{K^2}{2}qA \times \nonumber \\
&&\qquad  \times \frac{4\pi\sigma_1t(q)/\varepsilon_sV_0}
  {[1+4\pi\sigma_2t(q)/\varepsilon_sV_0]^2+[4\pi\sigma_1t(q)/\varepsilon_sV_0]^2},
    \, 
    \\
&&A = 8b(q)(\varepsilon_1 +\varepsilon_0)
\varepsilon_0^2 \varepsilon_s
\exp [-2q(a+d)],\text{  }  \,   \nonumber
\\
\label{eq:V}
&&\frac{\Delta V}{V_0}=\frac{K^2}{2}A    \frac{1+4\pi\sigma_2t(q)/\varepsilon_sV_0}
  {[1+4\pi\sigma_2t(q)/\varepsilon_sV_0]^2+[4\pi\sigma_1t(q)/\varepsilon_sV_0]^2}, \nonumber
\end{eqnarray}
\begin{eqnarray}
b(q)=(b_1(q)[b_2(q)-b_3(q)])^{-1}
  \, , \nonumber  \end{eqnarray}
\begin{eqnarray}
t(q)=[b_2(q)-b_3(q)]/2b_1(q)\, , \nonumber \end{eqnarray}
\begin{eqnarray}
&&b_1(q)=(\varepsilon_1+\varepsilon_0)(\varepsilon_s+\varepsilon_0)
- (\varepsilon_1-\varepsilon_0)
(\varepsilon_s-\varepsilon_0)e^{-2qa}\, ,  \nonumber
\\
&&b_2(q)=(\varepsilon_1+\varepsilon_0)(\varepsilon_s+\varepsilon_0)
+ (\varepsilon_1+\varepsilon_0)
(\varepsilon_s-\varepsilon_0)e^{-2qd}\, ,  \nonumber
\\
&&b_3(q)=
(\varepsilon_1-\varepsilon_0)(\varepsilon_s-\varepsilon_0)e^{-2qa}
+ \, \nonumber \\
&&\quad ~~~~~~~~~~~~~~~~~~~~~~~~~~ + (\varepsilon_1-\varepsilon_0)
(\varepsilon_s+\varepsilon_0)e^{-2q(a+d)}  \, \nonumber,
\end{eqnarray}

where $K^2$ is the coefficients of the electromechanical
coupling for lithium niobate; $q$ and $V_0$ are the wavevector
and SAW velocity, respectively; $a$ is the gap
between the piezoelectric lithium niobate substrate
and the sample; $d$ is the depth of the location of the
conducting channel (which is determined by technologists); $\varepsilon_1$, $\varepsilon_0$ and $\varepsilon_s$ are the dielectric constants for
lithium niobate, vacuum, and the sample, respectively;
and $b$ and $t$ are complex functions depending on
$a$, $d$, $\varepsilon_1$, $\varepsilon_0$ and $\varepsilon_s$. The resultant dependence is shown
in Fig.~\ref{Fig5_X2017scaling}.

\begin{figure}[t]
\centerline{
\includegraphics[width=7.9cm,clip=]{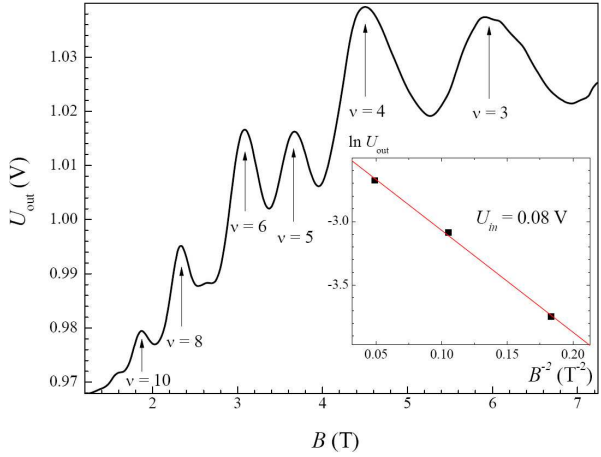}
} \caption{Dependence $U_{out} (B)$ for frequency $f$=1130 MHz at
$T$= 4.2; $\nu$ is the filling factor; the inset shows the
dependence of $\ln U_{out}$ от $B^{-2}$.
\label{Fig6_X2017scaling}}
\end{figure}

The microwave measurements on sample no. 1
were made in magnetic fields up to 8 T in the frequency
interval 100-1200 MHz in the temperature
range (1.7–4.2) K. Figure~\ref{Fig6_X2017scaling} shows the dependence of
signal amplitude $U_{out}$ at the outlet of the coplanar
waveguide on the magnetic field ($f$=1130~MHz, $T$=4.2 K). Analogously to dependences $\Delta\Gamma (B)$ and
$\Delta V(B)/V(0)$ measured using the acoustic technique,
first the Shubnikov–de Haas oscillations appear on
the field dependence of $U_{out}$ (up to 3 T), followed by
oscillations in the IQHE regime (in fields above 3 T).

The expression for calculating the real component
$\sigma_1$ of the ac conduction in the microwave method has
the form~\cite{Engel1993,Endo}
\begin{eqnarray}
\sigma_1=-\frac{w}{Z_0 l} \ln \left[\frac{U_{out}}{U_{in}}\right]
\sqrt{1+\left[\frac{v_{ph}}{l \omega}
\ln\left(\frac{U_{out}}{U_{in}}\right)\right]^2},
\label{s1CPW}
\end{eqnarray}
where $U_{in}$ is the signal amplitude at the input of the
coplanar waveguide, $Z_0$=50~$\Omega$ is the characteristic
impedance of the coplanar waveguide (without a sample),
$l$=5.3~cm is the length of the signal (“serpentine”)
wire of the coplanar waveguide, $w$=26~$\mu$m is
the width of the gap between the signal wire and
earthed conductor, $v_{ph}=c
\sqrt{2/(1+\varepsilon_{\text{GaAs}})}$=1.14$\times$10$^8$~m/s is the phase velocity of the wave passing
through the coplanar waveguide, and $\varepsilon_{\text{GaAs}}$=12.9 is the dielectric constant of the i-GaAs substrate on
which the coplanar waveguide has been formed.

It can be seen that for calculating $\sigma_1$, the value of
$U_{in}$ is required. It is extremely difficult to measure this
quantity; for this reason, we have used the following
assumptions for determining $U_{in}$.

(1) The signal at the output of the coplanar waveguide
is given by $U_{out}=U_S (B)
+ U_L$, where $U_S$ is the signal determined by the conductance
of the sample and $U_L$ is the background signal characterizing
the leakage through the coplanar waveguide.
Our measurements have shown that the change in the
phase of signal $U_{out}$ in a magnetic field is 20$^\circ$-50$^\circ$.
With such a change in the phase, the geometrical (vector)
sum of signals $U_S$ and $U_L$ is very close to their algebraic
sum. This enabled us to write signal $U_{out}$ in the
form of the above-mentioned algebraic sum.

(2) In zero magnetic field, useful signal $U_S$ is equal
to zero because the electric field of a quasi-TEM wave
is screened by electrons almost completely for $B$=0. A
noticeable $U_S$ signal appears only after the application
of a magnetic field, when the conductance of the sample
decreases.

(3) Background signal $U_L$ (at a preset frequency
and temperature) does not change in a magnetic field
and only produces a constant background of the output
signal $U_{out}$. This is confirmed by experiments with
coplanar waveguides.

Thus, taking into account assumptions (1)–(3), we
could reduce the calculation procedure to the following
steps.

(i) Subtraction of the amplitude of background signal
UL from the output signal $U_{out}$.

(ii) Determination of $U_{in}$ from the dependence of
$\ln (U_S, \nu)$ on $B^{-2}$ (here $\nu$ is the filling factor for
the Landau levels). This procedure was based on the
following considerations.~\cite{Drichko2014} In the magnetic fields
corresponding to minima of IQHE oscillations, electrons
are known to be localized, and ac conductance is
of the hopping type. This fact is confirmed by the inset
to Fig.~\ref{Fig5_JETP2017scaling}, which shows the temperature dependence of
the ac conductance for filling factors 3 and 4.
The flattening of the curves upon cooling indicates
that the conduction mechanism becomes hopping
indeed. In this case (see, for example, review~[\onlinecite{Parshin}]),
$\sigma_1 \propto B^{-2}$ (to the logarithmic accuracy) in a strong
magnetic field. If we disregard for simplicity the radicand
in formula~\ref{s1CPW} (which is significant only at low
frequencies, $\omega/2\pi \leq 300$~MHz), this gives
\begin{eqnarray}
\sigma_1=-\frac{w}{Z_0 l} \ln \left[\frac{U_{S}}{U_{in}}\right]
=-\frac{w}{Z_0 l} \left[\ln (U_{S}) -\ln (U_{in})\right]
  \nonumber 
\end{eqnarray}
whence
\begin{eqnarray}
\ln (U_{S})=-\frac{Z_0 l}{w} \sigma_1 + \ln (U_{in}) .
  \nonumber 
\end{eqnarray}

Therefore, for $B\rightarrow \infty$, conductance $\sigma_1\rightarrow$0 and
$\ln (U_{S}) \rightarrow \ln (U_{in})$. The inset to Fig.~\ref{Fig6_JETP2017scaling} shows the dependence
of $\ln
(U_S, \nu)$ от $B^{-2}$ for the maxima of the output
signal or, in other words, for the conductance minima
corresponding to occupation numbers $\nu$= 4, 6, 8. The
intersection of the line plotted using the linear approximation
with the ordinate axis gives $\ln (U_{S},
\nu)$=–2.52,
which corresponds to $U_{in}$= 0.08~V.

\begin{figure}[t]
\centerline{
\includegraphics[width=7.9cm,clip=]{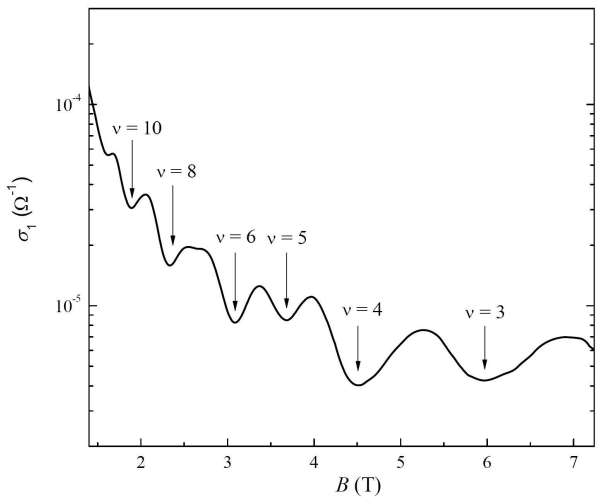}
} \caption{Dependence $\sigma_1 (B)$ for frequency $f$=1130 MHz at
$T$=4.2 K; $\nu$ is the filling factors.
\label{Fig7_X2017scaling}}
\end{figure}

(iii) The determination of the real component of
the high-frequency conductance by substituting the
obtained value of $(U_{in})$ into formula (\ref{s1CPW}). The result of
such a calculation is shown in Fig.~\ref{Fig7_X2017scaling}.

At the peaks of oscillations, $\sigma_1 (B)$ have an appreciable
value in all magnetic fields ($\sigma_1 >$10$^{-5}$$\Omega^{-1}$); this
corresponds to the conduction over extended states in
the vicinity of the centers of the Landau levels, which
is independent of frequency. It follows hence that the
peaks of conductance oscillations, which are measured
in the same magnetic field using different techniques,
should coincide; actually, however, a discrepancy
in these values is observed.

We think that this discrepancy is due to inaccuracy
in determining conductance $\sigma_1$ with the help of the
microwave method. The frequency independence of
the conductance at the maxima enabled us to compare
the results of the microwave and acoustic techniques.
To this end, each $\sigma_1 (B)$ dependence obtained by the
microwave method was multiplied by correcting coefficient
$K$, which was chosen so that the oscillation
maxima determined using the acoustic and microwave
techniques coincided. This precisely forms the additional
(after stages (i)–(iii)) procedure for processing
of the results.

\begin{figure}[t]
\centerline{
\includegraphics[width=7.9cm,clip=]{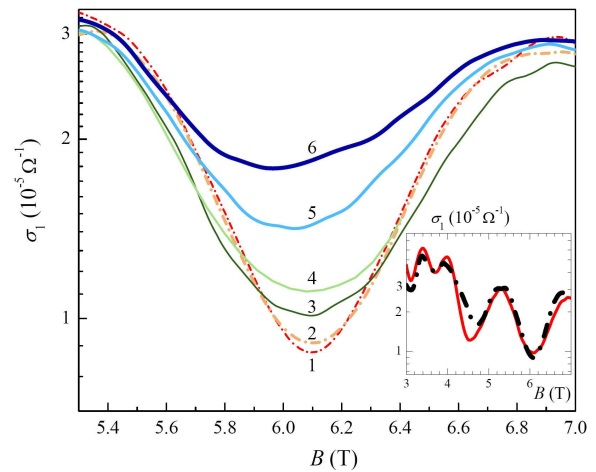}
} \caption{Dependences $\sigma_1 (B)$ for different frequencies:
(1) 30MHz and (2) 142 MHz (acoustic method);
(3) 250 MHz, (4) 328 MHz, (5) 800 MHz, and
(6) 1130 MHz (microwave technique); $T$ = 4.2 K. The
inset shows the $\sigma_1 (B)$ dependences obtained by superposition
of curves for $f\approx$220~MHz; solid and dashed curves
correspond to the microwave and acoustic techniques.
 \label{Fig8_X2017scaling}}
\end{figure}

The results of this processing for different frequencies
and $B$=6.1~T at $T$=4.2~K are shown in Fig.~\ref{Fig8_X2017scaling}. The
ac conductance for $B<$5.5~T is frequency-independent
to within the experimental error.

The inset to
Fig.\ref{Fig8_X2017scaling} confirms the correctness of the processing used:
the curves measured with the help of these different
methods at close frequencies in all magnetic fields
coincide to within the experimental error after the
superposition of the conductance peaks.

\begin{figure}[t]
\centerline{
\includegraphics[width=7.9cm,clip=]{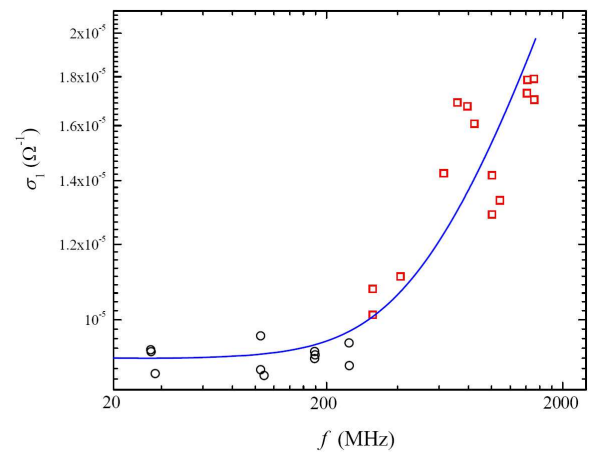}
} \caption{Dependences $\sigma_1 (f)$ at the minimum with filling factor  $\nu$=3 at $T$=4.2 K; $\bigcirc$ are the data
obtained using the acoustic method; $\square$ correspond to
the microwave technique. The solid scaling curve $\color{blue}\textbf{---}$ is drawn
in accordance with the theory described in Section 4.
\label{Fig9_X2017scaling}}
\end{figure}

Figure~\ref{Fig9_X2017scaling} shows the frequency dependence of ac
conductance $\sigma_1$ at the minimum corresponding to
filling factor $\nu$=3 at $T$=4.2 K. It can be seen
that for $f \leq$200~MHz, the frequency dependence is
weak and $\sigma_1 (\omega) \approx
\sigma_{\text{DC}}$. $f$$\geq$200 MHz, the behavior
of the frequency dependence of σ1 changes significantly.
The interpretation of these results is given in
the next section.

\section{Discussion}
\subsection{Absorption Mechanisms and Theoretical Models}

In the range of high frequencies, the ac hopping
conduction is determined by the jump between the
states belonging to the so-called close pairs in which
the distances between the sites is much smaller than
the mean distance between the sites. Therefore,
absorption at high frequencies is determined by two site
clusters. The distance between the lower energy
levels $E$ in such pairs is determined by the combination
of the difference of one-site energies, $\varphi=\varphi_1 - \varphi_2$, and
the tunnel overlap integral $\Lambda(r)=\Lambda_0
e^{-r/\alpha}$ (where $\alpha$ is
the electron state localization length):
\begin{eqnarray}
E=\sqrt{\varphi^2 + \Lambda^2}.
  \nonumber 
\end{eqnarray}

There are two competing mechanisms of absorption,
viz., the resonant and relaxation mechanisms
(see, for example, [\onlinecite{Parshin}]). In the former case, the wave
attenuates due to direct absorption of ac quanta with $\hbar \omega
= E$. This mechanism is important at high frequencies
beyond our experimental frequency range.

The other (relaxation) mechanism is associated
with modulation of occupancies of sites by the electric
field of the wave, which varies with time. It noticeably
depends on the population relaxation time, which can
be represented in the form~\cite{Parshin}
\begin{eqnarray}
\frac{1}{\tau (E,r)}=\frac{1}{\tau_{\text{min}} (E)}\left(\frac{\Lambda(r)}{E}\right)^2.
 \nonumber 
 \end{eqnarray}

Here, $\tau_{\text{min}} (E)$ has the meaning of the minimal relaxation
time for pairs with separation $E$ between energy
levels. Power $Q$ absorbed by a pair can be written in the
form
\begin{eqnarray}
Q = \frac{1}{2}
|e \textbf{E}_0 \textbf{r} |^2
\frac{\omega^2 \tau} {1+(\omega \tau)^2}
\left(-\frac{\partial n_o}{\partial E}\right).
 \nonumber 
 \end{eqnarray}

The energy absorbed by a pair depends on two
parameters - $E$ and $r$ (via $\Lambda(r) \propto e^{-r/\alpha}$)), and
the total absorbed power is determined by the sum of
the contributions from individual pairs. This summation
can be expressed as the integral with respect to these parameters, which is evaluated taking into
account their distribution functions. It turns out that
pairs with $E \lesssim kT$ are important in this case. The result
of summation over the pairs with different distances
between the sites is determined by the exponential
spread of tunnel integrals $\Lambda(r) \propto e^{-r/\alpha}$ in the case of a
smooth distribution of distances between the centers.
The final result depends on product $\omega \tau_{\text{min}}
(kT)$. In the
most interesting case for our analysis, when
\begin{eqnarray}
\omega \tau_{\text{min}} (kT) \ll 1
 \nonumber 
 \end{eqnarray}
the pairs with $\tau (kT)$$\approx$
$\omega^{-1}$ (i.e., those in which the distributions
of occupancies can reach equilibrium over a
time on the order of the ac wave period) are most
important. The characteristic size of such pairs is
\begin{eqnarray}
r_{\omega}=\alpha \ln [\Lambda_o/kT \sqrt{\omega \tau_{\text{min}} (kT)} ] \gg \alpha.
 \nonumber 
 \end{eqnarray}

For a 3D sample, the real part of the conductance
turns out to be proportional to $\omega r_{\omega}^3$, which is usually
expressed in the form $\sigma_1 \propto \omega^s$, where exponent $s$ is several
times smaller than unity and weakly depends on
temperature.\cite{Shklovskii}

It can be seen that upon a decrease in the frequency,
the size of a characteristic pair increases, and
pairs at low frequencies overlap to form large clusters
(the so-called multiple hopping regime takes place).
The size of a characteristic cluster in this case is determined
by the stabilization of the population of sites
during half-period $\pi/\omega$ of the wave. Therefore, the
characteristic size of a cluster increases upon a
decrease in frequency. With a further decrease in frequency,
clusters merge into an infinitely large cluster
responsible for static conduction.\cite{Shklovskii2,Dyre,Zvyagin}

There are several alternative models for describing
the crossover from the ac conduction to dc conduction;
the results of these models turn out to be close. In
this study, we are using the Zvyagin model~\cite{Zvyagin} formulated
for the situation in which the spread in the rates
of transitions between the centers is exponentially
wide.

In this model, transitions between localized states $\lambda$
and $\lambda'$ are characterized by the rates
\begin{eqnarray}
\Gamma_{\lambda\lambda'}=\Gamma_0 e^{-\eta_{\lambda\lambda'}},
 \nonumber 
 \end{eqnarray}
 where
 \begin{eqnarray}
\eta_{\lambda\lambda'} = \frac{2r_{\lambda\lambda'}}{\alpha}+
\frac{| \varepsilon_{\lambda}-\varepsilon_F|+
| \varepsilon_{\lambda'}-\varepsilon_F|+
| \varepsilon_{\lambda}-\varepsilon_{\lambda'}|}
{2kT}.
 \nonumber 
 \end{eqnarray}

Like in the static problem on percolation conduction,
two sites for the chosen value of $\Gamma$ are assumed to
be coupled if
\begin{eqnarray}
\Gamma_{\lambda\lambda'} > \Gamma.
 \nonumber 
 \end{eqnarray}

 For an exponentially broad distribution of $\Gamma_{\lambda\lambda'}$, the
static conductance is determined by a certain critical
value $\Gamma_c$, which corresponds to the formation of an
infinite cluster.

Let us now choose a finite cluster $k$ in which the
minimal transition rate satisfies the condition $\Gamma_k^* > \Gamma > \Gamma_c$. For a wide spread in transition rates, it is $\Gamma_k^*$
that determines the time of stabilization of equilibrium
in the $k$th cluster. Therefore, quasi-equilibrium is
established in the cluster if
\begin{eqnarray}
\tau(\Gamma_k^*) < \omega^{-1}.
 \nonumber 
 \end{eqnarray}

For $\Gamma > \Gamma_c$, all clusters in the system are finite.

Let us now define quantity $\Gamma_{\omega}$ by the condition
$\tau(\Gamma_{\omega})$=$1/\omega$; in this case, all clusters are quasi-equilibrium,
and relaxation losses occur only due to transitions
between such clusters. The main idea is that for a
wide spread in the transition rates, we can consider
only the “bottleneck,” i.e., the transition at the lowest
rate, which separates two quasi-equilibrium parts of a
large cluster. This transition divides the cluster into
two parts; as a result, a structure resembling a two-site
system is formed; the sites of this system consist of
parts of the initial cluster, in which transitions occur
with relatively high rates. These parts play the role of
the renormalized sites of the two-site model. Summation
over such clusters is carried out on the basis of the
known statistics of clusters in percolation systems~\cite{Shklovskii2,Stauffer} taking into account the fact that clusters with a
relaxation time of occupancies on the order of $\omega^{-1}$ are
important.

Since the proposed procedure can be reduced to
the renormalization of the properties of effective sites,
it is not surprising that the theory predicts universal
frequency dependences of ac conductance. According
to [\onlinecite{Zvyagin}], the result for the 3D case can be written in the
form
\begin{eqnarray}
\frac{\sigma_1(\omega)}{\sigma_{\text{DC}}} \ln ^\xi
\left( \frac{\sigma_1(\omega)}{\sigma_{\text{DC}}} \right) =
\frac{\omega}{\omega_S}.
 \label{s1ksi}
\end{eqnarray}
where $\xi$ and
\begin{eqnarray}
\omega_S=\frac{\alpha}{C_0 \eta_c^{\xi-1}} \sigma_{\text{DC}}.
 \nonumber 
\end{eqnarray}
are the scaling parameters that generally depend on
temperature. In the above expressions, $C_0$ is the capacitance
per site and $\eta_c$ is the percolation threshold.

The derivation given in [\onlinecite{Zvyagin}] does not contain specific
features of 3D systems except that other critical
indices appear in the 2D case. In addition, the dimension
of the conductivity in the 2D case coincides with
the dimension of the total conductivity. For our purpose,
these differences are not very significant,
because we analyze only the frequency dependences of
the hopping ac conductance in the regime of the integer
quantum Hall effect, treating critical index $\xi$ and
frequency $\omega_S$ as fitting parameters. We do not consider
here the complicated and interesting question concerning
factor $\sigma_{\text{DC}}$ in the IQHE regime,~\cite{Polyakov} which
affects frequency $\omega_S$.

The above scaling relations will be used below for
analyzing the ac conductance at the oscillation minima,
which corresponds to the regime of the integer
quantum Hall effect, where the conduction is of the
hopping type. Indeed, at low temperatures (1.7~K), the
hopping nature of the ac conductance at the minima
of oscillations is beyond any doubt, because the temperature
dependence is flattened (see inset to Fig.\ref{Fig5_X2017scaling}),
and an increasing frequency dependence of $\sigma_1$ is observed even for $f>$100~MHz. However, at higher
temperatures (4.2 K) and low frequencies, the ac conductance
is close to the static ac conductance, which
depends on temperature in the activation manner (this
is illustrated by straight lines in Fig.~\ref{Fig5_X2017scaling}) because large
clusters play the major role in this approximation. The
frequency dependence corresponds to the typical
behavior of hopping ac conductance only for $f>$~200~MHz.~\cite{Drichko2000}

The curve plotted in Fig.~\ref{Fig9_X2017scaling} for sample no. 1 ($T$=4.2 K) corresponded to scaling relation (3) for parameters
$\omega_S$=4.0$\times$10$^9$~s$^{-1}$ and $\xi$=0.5. The value of $\sigma_{\text{DC}}$=9.1$\times$10$^{-6}$~$\Omega ^{-1}$ was obtained by extrapolating the frequency
dependence of $\sigma_1$ in the frequency range $f$$\leq$200~MHz.

\begin{figure}[t]
\centerline{
\includegraphics[width=7.9cm,clip=]{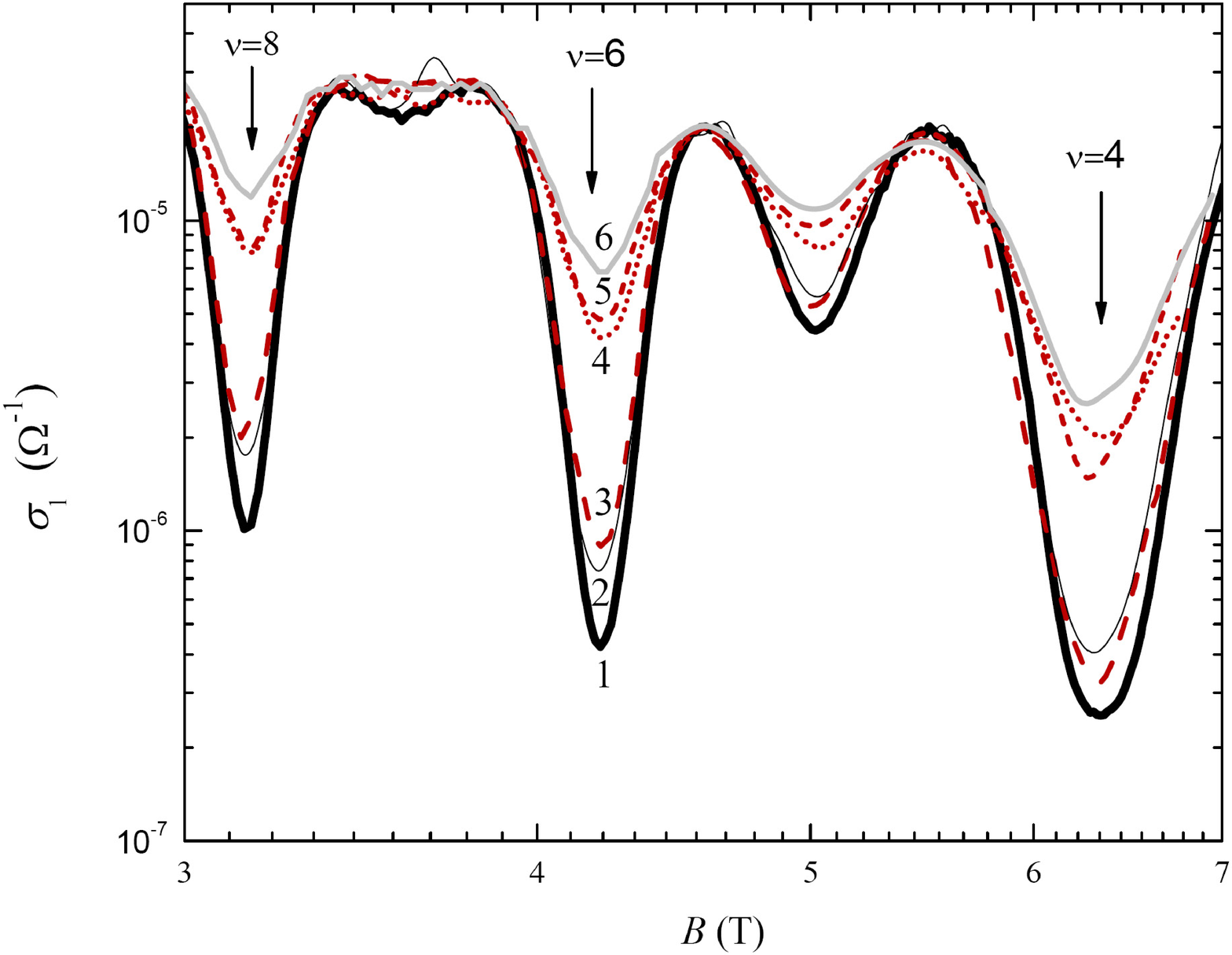}
} \caption{Dependences of $\sigma_1$ on $В$ at different frequencies $f$,
MHz: (1) 30, (3) 197 (acoustic method); (2) 148, (4) 585,
(5) 906, and (6) 1191 (microwave technique); $T$=1.7 K;
sample no. 2.
\label{Fig10_X2017scaling}}
\end{figure}

It is convenient to trace the magnetic-field dependence
of scaling parameters on sample no. 2, in which
hopping conduction was observed at $T$=1.7 K for several
values of the magnetic field. Figure~\ref{Fig10_X2017scaling} shows the
magnetic-field dependence of the conductance of this
sample for different frequencies.

\begin{figure}[t]
\centerline{
\includegraphics[width=7.9cm,clip=]{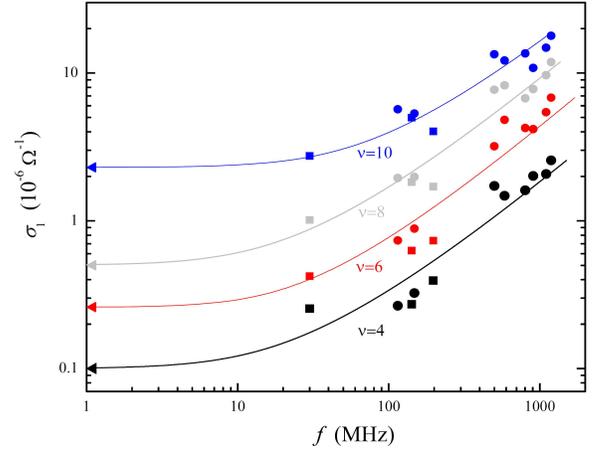}
} \caption{Dependences $\sigma_1 (f)$ at the minima of oscillations
with $\nu$ = 4, 6, 8, 10. Squares correspond to acoustic
method, circles are the data obtained by microwave technique,
and triangles correspond to $\sigma_{\text{DС}}$, $T$=1.7~K.
\label{Fig11_X2017scaling}}
\end{figure}

The frequency dependence of the conductance of
sample no. 2 in different magnetic fields is shown in Fig.~\ref{Fig11_X2017scaling}. For plotting the curves in Fig.~\ref{Fig11_X2017scaling}, we selected
two parameters (scaling frequency $\omega_S$=$2\pi f_S$ and dimensionless
scaling parameter $\xi$). We used the value of $\sigma_{\text{DС}}$,
which was determined for this sample in [\onlinecite{Kanel2}] at $T$=1.7 K.

The resultant values of the scaling parameters for
sample no. 2 are given in Table~\ref{scalingtab}.

\begin{table}[h]\begin{center}\caption{Scaling parameters.}
\vspace{0.5cm}
\begin{tabular}{|c|c|c|c|c|}
     \hline
$\nu$ & $B$ (T) & $\sigma_{\text{DC}}$($\Omega ^{-1}$) & $f_S$ (MHz) & $\xi$ \\
\hline
4 & 6.3 & 1E-7 &  25 & 0.7  \\    
6 & 4.22 & 2.6E-7  & 32 &  0.6 \\ 
8 & 3.16 & 5E-7 &  26 &  0.7\\ 
10 & 2.53 & 2.3E-6  & 88 &  0.7  \\ %
\hline
\end{tabular}\label{scalingtab}
\end{center}
\end{table}

It can be seen from Table~\ref{scalingtab} that the scaling parameters
indeed depend on the magnetic field.

\begin{figure}[t]
\centerline{
\includegraphics[width=7.9cm,clip=]{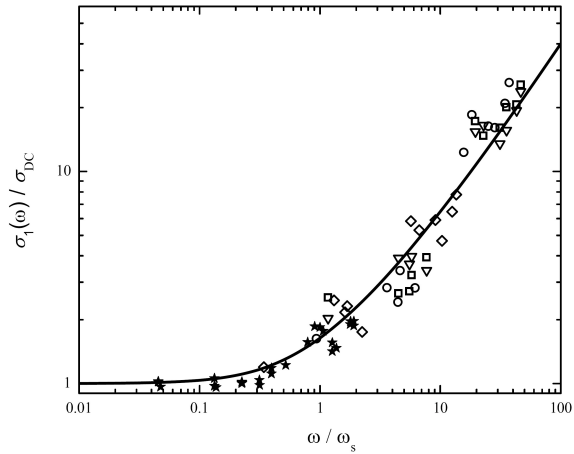}
} \caption{Dependence of $\sigma_1 / \sigma_{\text{DC}}$ on $\omega / \omega_S$: $\bigstar$ correspond
to sample no. 1 ($T$=4.2 K), $\square$ correspond to sample
no. 2, $\nu$=4; $\bigcirc$, sample no. 2, $\nu$=6; $\bigtriangledown$, sample
no. 2, $\nu$=8; $\diamondsuit$, sample no. 2, $\nu$=10.
\label{Fig12_X2017scaling}}
\end{figure}

Scaling equation~\ref{s1ksi} can be used to express explicitly
the dependence of ratio $\frac{\sigma_1}{\sigma_{\text{DC}}}$ on
$\frac{\omega}{\omega_S}$:
\begin{equation}
\frac{\sigma_1(\omega)}{\sigma_{\text{DC}}}=\frac{\omega}{\omega_S} \left[\xi W
\left(\frac{1}{\xi}\left[\frac{\omega}{\omega_S}\right]^{1/\xi}\right)
\right]^{-\xi},
\label{s1sdc}
\end{equation}
where $W (x)$ is the Lambert function. This dependence
is plotted in Fig.~\ref{Fig12_X2017scaling}.

It can be seen from the figure that all points for two
different samples, different temperatures, and magnetic
fields lie on the same curve, which proves the
universality of the frequency dependence of the conductance
in the hopping conduction conditions in this
model.

It turned out that the value of $\xi$ weakly affects the
frequency dependence: the change in $\xi$ from 0.4 to 0.7
leads to satisfactory agreement with experimental data
at all oscillation minima (the curve in Fig.~\ref{Fig12_X2017scaling} was plotted
for $\xi$=0.7). Since $\xi$ is a combination of critical
indices, we can rightfully assume that this quantity
weakly depends on the magnetic field. We have not
detected such a dependence within the error of our
experiment.

\section{Conclusions}

The high-frequency hopping conduction observed
in magnetic fields corresponding to the conductance
minima in the regime of the integer quantum Hall
effect was measured in p-GeSi/Ge/GeSi structures
with quantum wells in a wide frequency range (30-1200 MHz). The measured frequency dependences
are in agreement with the model presuming jumps
between localized electron states belonging to isolated
clusters. At high frequencies, predominant clusters are
pairs of close states; upon a decrease in frequency,
large clusters play the major role; as the frequency
tends to zero, such clusters merge into an infinitely
large percolation cluster. In this case, the frequency
dependences of the ac conductance can be represented
by a single universal curve (scaling). The scaling
parameters and their magnetic-field dependence
have been determined.

\begin{acknowledgements}

The work of one of the authors (I.L.D.) was supported
by the Presidium of the Russian Academy of
Sciences.

\end{acknowledgements}


%

\end{document}